\documentclass[final]{elsarticle}

\usepackage{hyperref}

\usepackage{graphicx}
\usepackage{caption}
\usepackage{subcaption}
\usepackage{amssymb}

\usepackage{xcolor}
\usepackage{float}
\usepackage{listings}

\newfloat{lstfloat}{htp}{lop}
\floatname{lstfloat}{Listing}

\journal{Journal of Computational Physics}

\begin{document}

\begin{frontmatter}

\title{Simulating the Injection of Magnetized Plasma without Electromagnetic Precursor Wave}

\author[nwu]{Patrick Kilian}
\cortext[mycorrespondingauthor]{Corresponding author}
\ead{28233530@nwu.ac.za}

\author[nwu]{Felix Spanier}

\address[nwu]{Centre for Space Research, North-West University, Potchefstroom, South Africa}


\begin{keyword}
Particle-in-Cell \sep Magnetized \sep Collisionless Plasma \sep Electromagnetic Simulation \sep Boundary Condition
\end{keyword}

\end{frontmatter}


\section{Introduction}

This note aims to explain how to inject magnetized plasma through an open
boundary into the simulation domain of a particle-in-cell simulation. If the
magnetic field at the boundary is constant in time, i.e., if magnetized plasma
of constant magnetization is injected at a steady rate, this does not present
any challenges beyond injecting the particles at a fixed rate and possibly
absorbing plasma waves impinging on the wall. If, however, the magnetization or
the injection rate changes, a time-varying magnetic field is present.
The classical use case for this scenario is a shock front moving through
a plasma into the simulation volume.

The time variation in magnetic field obviously produces a curl of the electric
field. This new electric field in turn produces a magnetic field of its own.
Or in other words, an electromagnetic wave is launched. This effect might be
desired in the implementation of antennas that launch electromagnetic radiation
into the simulation domain. When injecting magnetized plasma this high
frequency effect is undesired and might -- depending on the amplitude of the wave -- produce
unphysical or at least undesired effects in the medium ahead of the injected
material.  From observations \cite{Dryer_1974,Howard_1982,Schwartz_1988}
 it is known that shock waves do not continouosly emit
electromagnetic radiation ahead of the shock front, which is what should be
reproduced in sumulations. The goal is, therefore, to find an injection method
that is capable of handling time variable magnetic fields at the injection
site without launching an electromagnetic percursor wave.

Both antennas and injection of magnetized plasma is often performed by
prescribing a current density instead of directly setting the magnetic field.
This has the advantage that the resulting magnetic field is automatically
divergence free as it is calculated from Ampere's Law.

One example of using a current pulse to push magnetized plasma is given in
\cite{Lembege_1987}. The authors there also note that the shape of the current
pulse is not arbitrary, but has to be chosen carefully to avoid the generation
of a precursor wave. Their choice for the injection profile was a cosine half
wave switching smoothly from zero to a desired value and subsequently holding
that value constant. Two different period lengths of the cosine are discussed,
$8~\omega_\mathrm{p,e}^{-1}$ and $60~\omega_\mathrm{p,e}^{-1}$, where $\omega_\mathrm{p,e}$ is the electron plasmafrequency at the initial, uncompressed density. The first case
works well as a driver, but still launches a visible electromagnetic wave. The
slower injection of the latter case avoids this effect. Additionally, the
current is not injected in a single location, but over a finite spatial range
of 14 Debye lengths, which helps to reduce high $k$ noise, equivalent to high
frequency waves.

Another example of a time variable magnetic field that passes the boundaries of
the simulation domain is presented in \cite{Hurtig_2003}. In that case, a
non-homogeneous magnetic field is considered, combined with a moving simulation
domain that tracks a bunch of particles. This application, however, uses an
electrostatic plasma model, that is free from electromagnetic waves by
construction.

In many other cases only unmagnetized plasma is injected and magnetic fields
are only present in the simulation domain through the self-consistent
interaction with the ambient plasma. Typical examples can be found in
\cite{Ardaneh_2016} and references therein, especially \cite{Nishikawa_2003}.

All the tests of the injection method presented below in section
\ref{sec:method} were performed with the electromagnetic particle-in-cell code
ACRONYM \cite{Kilian_2012}. It is a fully-parallelized code for the simulation
of collisionless plasma phenomena using standard algorithms of second order
accuracy in space and time. Electromagnetic fields are stored in a standard Yee
(\cite{Yee_1966}) grid. Particles are updated using the Boris push
\cite{Boris_1970,Penn_2003}. The current density that results from the
movements of the charge particles is deposited onto the grid using the method
of Esirkepov \cite{Esirkepov_2001}. This deposition, as well as the
interpolation of the electromagnetic fields to the particle position, use the
second order interpolation using a TSC shape function. The code allows to use
any of the different FDTD schemes listed in \cite{Vay_2011} to calculate the
update of the electromagnetic fields. However, for the purpose of this note,
only the standard second-order scheme is used, that uses a straightforward
approximation of the curl using four terms in the expression for central differences. The
open boundary through which the magnetized plasma is injected discards any
particles that might reach it from the simulation domain. All incoming waves
are absorbed by a perfectly matched layer (PML, see \cite{Berenger_1994}). More
precisely, a PML with a complex frequency shift is used, that is implemented
using time domain equations following the convolutional PML scheme described in
\cite{Berenger_2007}.

\section{Method}
\label{sec:method}

\begin{figure}
    \centering
    \begin{subfigure}[b]{0.45\textwidth}
        \includegraphics[width=\textwidth]{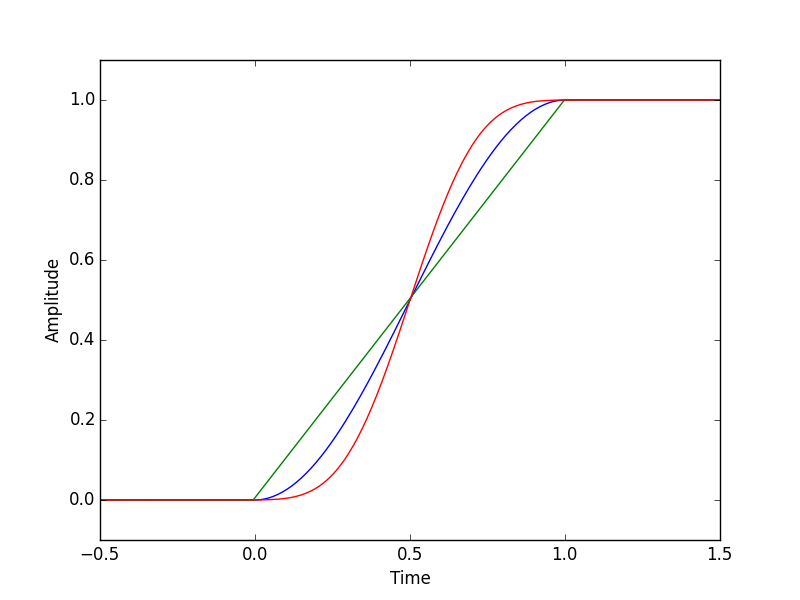}
        \caption{Step Response}
        \label{fig:step}
    \end{subfigure}
    \begin{subfigure}[b]{0.45\textwidth}
        \includegraphics[width=\textwidth]{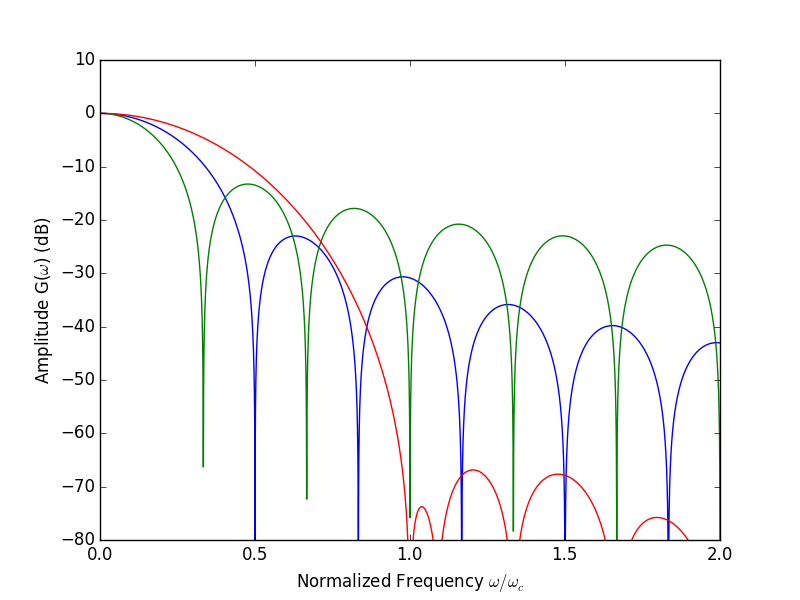}
        \caption{Spectral Content}
        \label{fig:dB}
    \end{subfigure}
    \caption{Amplitude over time (step response) and over frequency (frequency spectrum) of different window functions. A linear ramp, that result from passing the injection through a moving average filter and that injects at constant rate is shown in green. The sinusoidal half wave suggested in \cite{Lembege_1987} is shown in blue. The Blackman-Harris window suggested in this work is indicated in red.}
    \label{fig:windows}
\end{figure}

The electric field of the precursor wave is produced by the change in time of the
magnetic field. An obvious approach to remove the precursor is thus to feed the
injected magnetic field through a low pass filter to remove the steep rise. In
more technical terms, the derivative of the magnetic field, is convolved with a
digital low pass filter and then running sum of the resulting signal is
applied on the boundary. Figure~\ref{fig:step} shows the resulting
amplitude of the supplied magnetic field at the boundary of the simulation
domain as a function of time.

To see the effective reduction of high frequency components that could
propagate through the plasma, it is useful to also plot the
frequency dependence of the transfer function of the low pass filter. This is
shown in figure~\ref{fig:dB}. Alternatively, this can be seen as the amplitude
of the waves launched at the boundary as a function of frequency.

The simplest digital low pass filter is the moving average filter, that splits
up the change in magnetic field equally over $M$ time steps. In
figure~\ref{fig:windows} this is indicated by a solid green line. The sharp
step function in magnetic field is replaced by a linear increase and, as the
plot over frequency shows, amplitudes are reduced at higher frequencies.

Slightly more complicated digital filters are also available. One possible
choice is the sinusoidal half wave that was also used by \cite{Lembege_1987},
indicated in red.

The preferred solution is the use of a minimum 3-term Blackman-Harris window. This filter has the strongest suppresion of high-frequency components above the cut-off frequency for any filter that inject over a number of steps $M$.
(see \cite{Harris_1978} for a review of suitable digital filters). As figure~\ref{fig:dB} shows, there is practically no
components above a critical frequency $\omega_\mathrm{c}$. This critical
frequency depends on the number of steps $M$ and their length $\Delta t$
through
\begin{equation}
	\omega_\mathrm{c} = \frac{12 \pi}{M \Delta t} \quad .
\end{equation}

For comparison, the critical frequency of the rectangular window of a moving
average filter is given by
\begin{equation}
	\omega_\mathrm{c} = \frac{4 \pi}{\left(M+1\right) \Delta t} \quad .
\end{equation}
Figure~\ref{fig:windows} uses identical values of $M$ for all windows, which
explain why the moving average filter has its critical frequency at about one
third of the critical frequency of the Blackman-Harris window. However, it
still shows much larger amplitude, as the maximum attenuation is much lower.

For the sinusoidal window, the critical frequency is
\begin{equation}
	\omega_\mathrm{c} = \frac{6 \pi}{M \Delta t} \quad ,
\end{equation}
but the damping of high frequencies is not much better than for the moving
average filter.

The crucial step now is to make sure that the critical frequency is below the
gyro frequency of electrons $\Omega_\mathrm{c,e}$. All waves with frequencies
above the critical frequency are suppressed by the injection method. All waves
with frequencies close to the gyro frequency are strongly damped through
interaction with gyrating electrons (see \cite{Gary_2004,Schreiner_2017}).

\begin{figure}
    \centering
    \begin{subfigure}[b]{0.45\textwidth}
        \includegraphics[width=\textwidth]{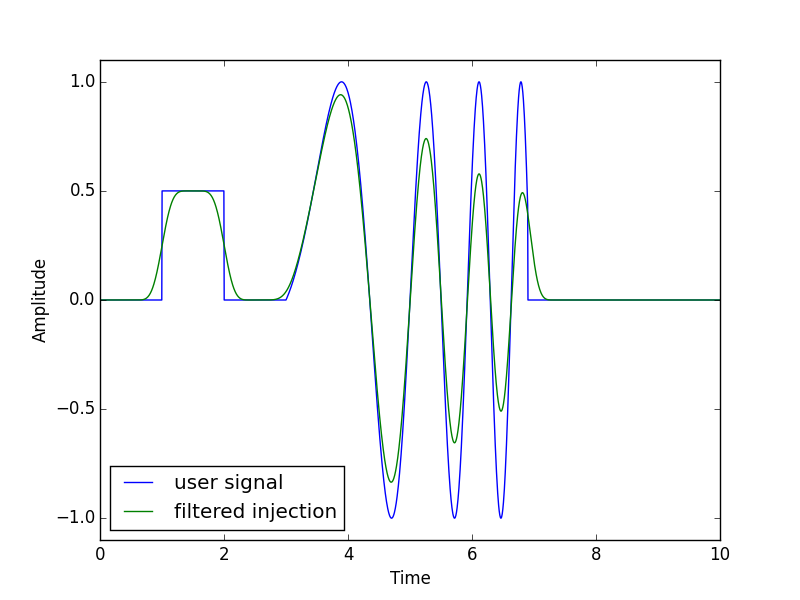}
        \caption{Wave form}
        \label{fig:signal}
    \end{subfigure}
    \begin{subfigure}[b]{0.45\textwidth}
        \includegraphics[width=\textwidth]{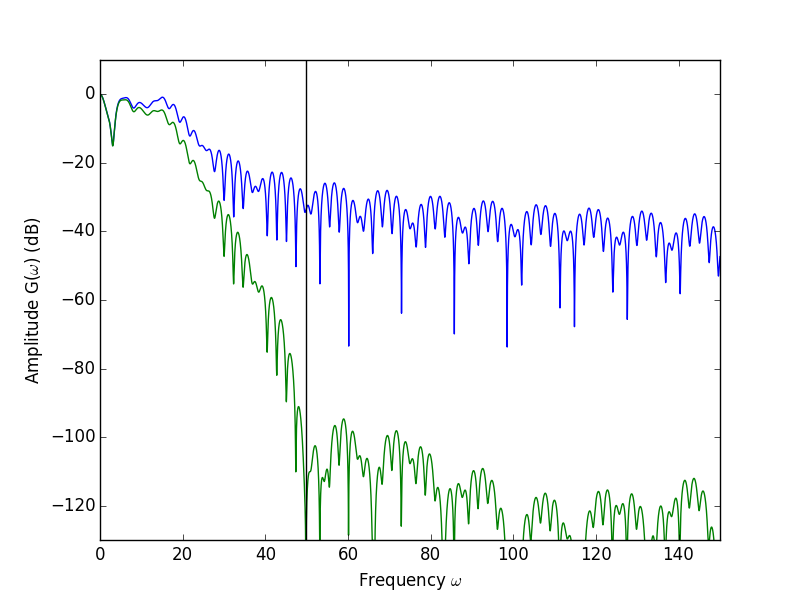}
        \caption{Spectral Content}
        \label{fig:spec}
    \end{subfigure}
    \caption{The left panel shows a possible signal that an user might select for injection and the actual low pass filtered field that would be injected. In the right panel the spectral content of the signal is indicated before and after the low pass filter. The selected cut-off frequency is indicated by a vertical line.}
    \label{fig:injection}
\end{figure}

To get even better results, it is advisable to inject the magnetic field not at
a single point, but through the entire thickness of the perfectly matched
layer. This removes the effect that the PML absorbs some of the injected
magnetic field, which makes injection of a well determined magnetic field hard.
Furthermore, it reduces the energy available to modes at large $k$ that might
propagate as high frequency waves into the simulation box.

Listing~\ref{code:example} below shows how the signal $f(t)$ supplied by the
user can be passed through the low pass filter to result in a filtered signal
$F(t)$ that contains no components above $\omega_c$.

\begin{lstfloat}
\begin{lstlisting}[language=Python,backgroundcolor=\color{lightgray},basicstyle=\footnotesize,numbers=left,xleftmargin=\parindent]
 # Step response of Blackman-Harris Window of length M
 W = numpy.cumsum(blackman(M))
 # Filtered output signal
 F = numpy.zeros(N)
 # "Circular" buffer
 b = numpy.zeros(M+1)
 for i in range(M//2, N):
     # find step
     delta = f(i*dt) - f((i-1)*dt)
     # write step response to buffer
     for m in range(M):
         b[m] += delta * W[m]
     b[M] += delta
     # add front of buffer to output
     F[i-M//2] += b[0]
     # cycle buffer
     for m in range(M):
         b[m] = b[m+1]
\end{lstlisting}
\caption{Possible implementation that decomposes the user supplied signal into steps that are convolved with the step responce of the window function that is used for low pass filtering. The infinitely long duration of the step response is implemented using a shifting buffer, that latches the last element.}
\label{code:example}
\end{lstfloat}

In the above implementation the Blackman-Harris window has to be normalized,
i.e., all $M$ entries have to sum to one. The described form is not the only
possible design. For example the code above implements a zero phase filter
without lag. If the user supplied function $f$ is not know ahead of time, it
might be necessary to switch to a causal, linear-phase filter that delays the
input signal by $M/2$ timesteps.

\subsection{Example case}

The main application that is driving the development of this method has been the
injection of magnetized shocks into a simulation box. Observations from e.g.
the Sun have shown that plasma shocks will travel through bulk plasma without
constantly emitting EM noise. It has also been shown that these plasma shocks
have a finite width.

A correctly modeled injection of a shock should show its propagation through
the medium without the emission of an EM emission when entering the box.
Unfortunately a comparison with other methods is difficult. The main reason
for this is that shock simulations have been performed in a very different
setup in the recent years: The generally used methods \cite{Gallant_1992} uses a
box with a wall on one side, where a plasma stream is reflected, which leads
to a subsequent shock front. This model is used, since the injection of a
magnetized plasma stream from one face of the box is not working well and
is typically too expensive.

Since a direct comparison of shock-like injection to other codes is not
possible, we will use a different benchmarking technique: Since the EM
noise moves away from the shock at the speed of light, the benchmark is
that the shock structure in the filtered and the unfiltered version
(after the EM pulse has moved away) will be unaltered.

To demonstrate the performance of the suggested injection method we have
simulated the injection of particles into a vacuum, initially without magnetic
field. After some time we switch on a magnetic field pointing perpendicular to
the flow direction. This setup is simulated with an instantaneous step function
as well as a filtered injection.

\begin{figure}
    \centering
    \begin{subfigure}[b]{0.45\textwidth}
        \includegraphics[width=\textwidth]{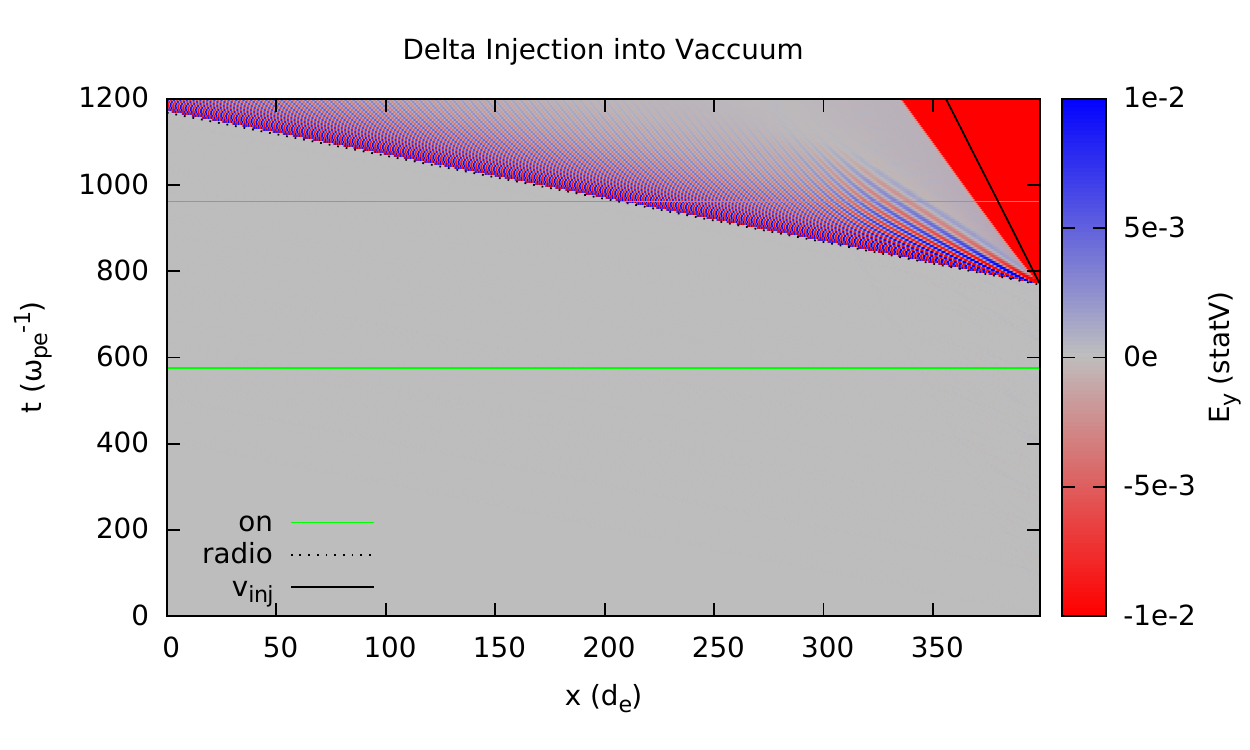}
        \caption{Step injection}
        \label{fig:pic_step}
    \end{subfigure}
    \begin{subfigure}[b]{0.45\textwidth}
        \includegraphics[width=\textwidth]{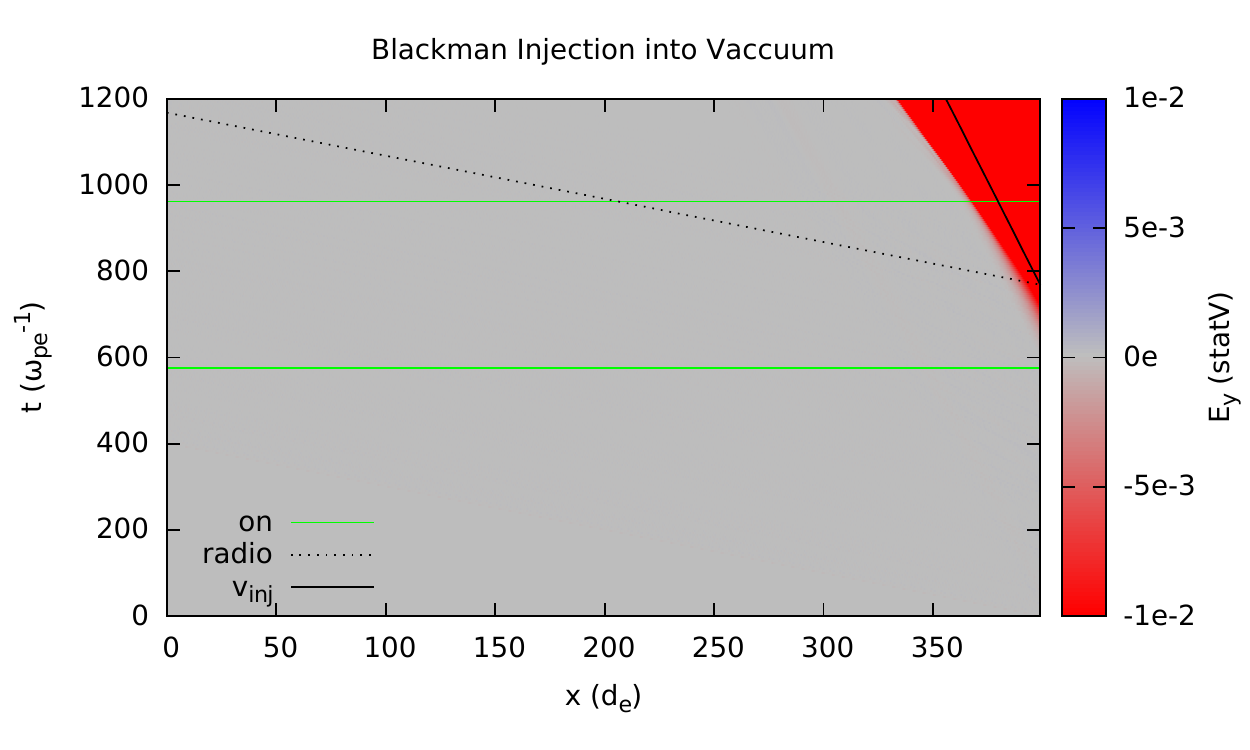}
        \caption{Filtered injection}
        \label{fig:pic_blackman}
    \end{subfigure}
    \caption{The left panel shows a transverse component of the electric field when changing the magnetization of incoming plasma with a sudden step function. Both the electromagnetic precursor wave and the large scale convective electric field are visible. In the right panel the injection is filtered with a Blackman-Harris filter, that cuts off components about the gyro frequency of electrons. This removes the undesired precursor wave.}
    \label{fig:pic}
\end{figure}

Figure \ref{fig:pic} shows the results in both cases. In the left panel
\ref{fig:pic_step} the injection with the sudden step is shown. The electric
field perpendicular to the flow direction and the magnetic field shows a clear
signature of a signal propagating at the speed of light. The amplitude of this
signal is $8\cdot10^{-2}\,\mathrm{statV}$, significantly larger than the
convective electric field caused by the $\vec{u} \times \vec{B}$ term of the
streaming magnetized plasma.

In the right panel \ref{fig:pic_blackman} the same situation is simulated
again, but with a filtered injection. The precursor waves is absent in this
case and only the convective electric field is visible. The noise level from
the finite number of computiational particles is $\pm
2\cdot10^{-4}\,\mathrm{statV}$ and even on that scale no precursor is
detectable. This shows that the digital low pass filter attenuates the
unphysical precursor wave by at least 50 dB.

\begin{figure}
    \centering
    \begin{subfigure}[b]{0.45\textwidth}
        \includegraphics[width=\textwidth]{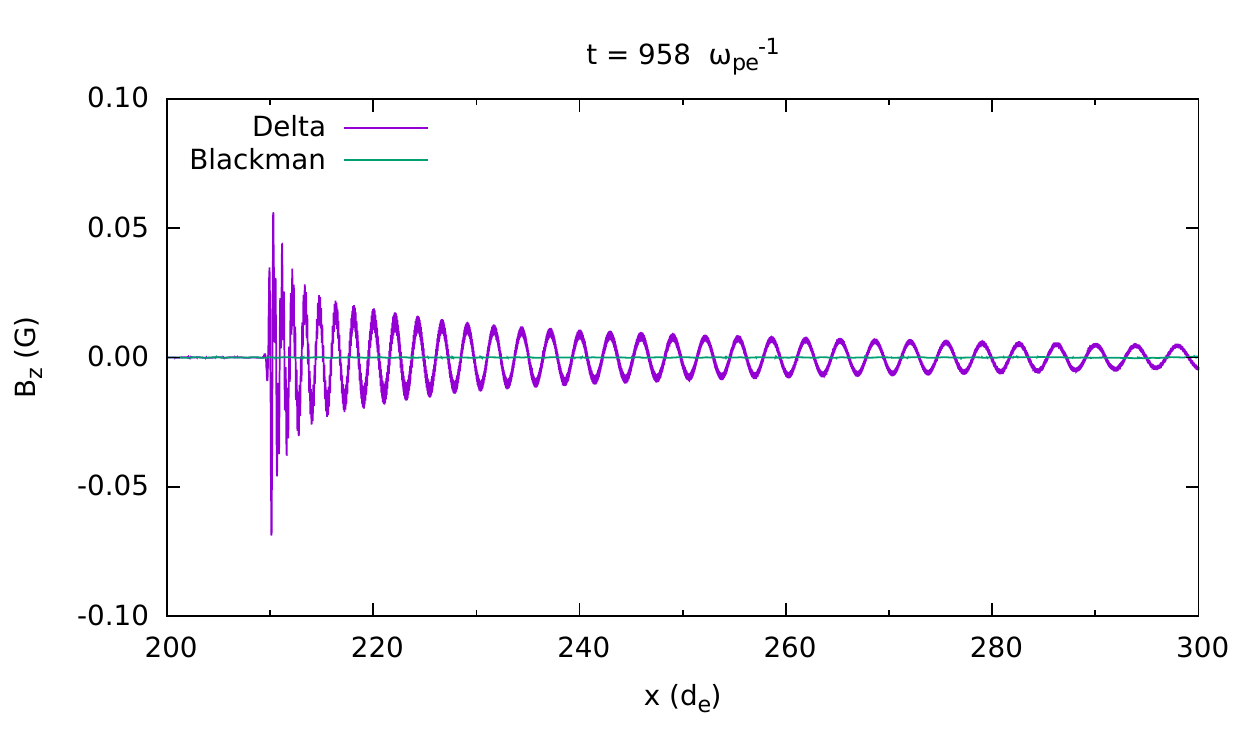}
        \caption{Electromagnetic precursor}
        \label{fig:pic_pulse}
    \end{subfigure}
    \begin{subfigure}[b]{0.45\textwidth}
        \includegraphics[width=\textwidth]{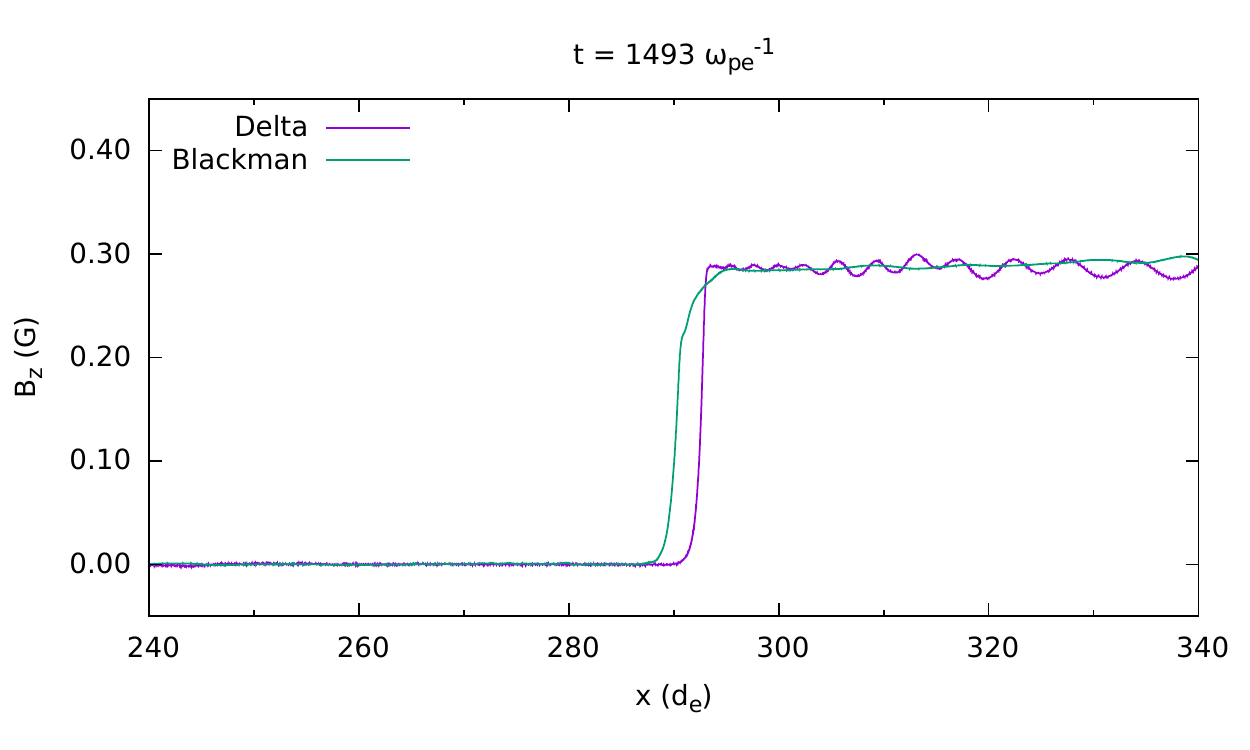}
        \caption{Shock profile}
        \label{fig:pic_front}
    \end{subfigure}
    \caption{The left panel shows the unphysical electromagentic precursor wave that is caused by the step injection. The use of the filtered injection removes the precursor. The right panel compares the resulting shock profile. There is a small difference in timing, but this is negligable for most application and could easily be adjusted for. The additional osciallations that are visible when using the step injection are physically not expected and show the same dispersive behaviour that is seen in the unphysical precursor.}
    \label{fig:profiles}
\end{figure}

Figure \ref{fig:profiles} shows two snapshot containing the precursor wave and
the desrired shock front and compare between the step injection and the
filtered injection.
It can be easily seen that the actual shock region is unaffected by the filtering
while the unphysical EM pulse is removed from the simulation domain.

In conclusion we report that standard digital low pass filters are suitable to
remove high frequency components from the wave form of the magnetic field
applied at a boundary. Selecting a cut-off frequency close to the gyro
frequency of the plasma leads to good results, without undesired precursor
waves. In principle different digital filters can be used, but the use of a
Blackman-Harris window offers very good performance at low numerical effort. If
necessary a Dolph-Chebyshev window would offer moderately better performance
but the filter coefficients are much harder to generate. Simpler windows
of a moving-average filter did not produce satisfactory results.

\section*{Acknowledgment}

The authors would like to thank Patricio Munoz for useful discussions and
feedback on an earlier version of this manuscript.
We acknowledge the use of the ACRONYM code supplied by its developers (Verein
zur F\"orderung kinetischer Plasmasimulationen e.V.). The authors gratefully
acknowledge the Gauss Centre for Supercomputing e.V. (www.gauss-centre.eu) for
funding this project (pr74se) by providing computing time on the GCS
Supercomputer SuperMUC at Leibniz Supercomputing Centre (www.lrz.de).
This work is based upon research supported by the National Research Foundation
and Department of Science and Technology. Any opinion, findings and conclusions
or recommendations expressed in this material are those of the authors and
therefore the NRF and DST do not accept any liability in regard thereto.

\section*{References}

\bibliography{references}

\end{document}